\documentclass[journal=jpccck,manuscript=article]{achemso}

\usepackage[version=3]{mhchem} 
\usepackage[T1]{fontenc}       

\author{Matteo Gerosa}
\email{mgerosa@uchicago.edu}
\affiliation{Department of Energy, Politecnico di Milano, via Ponzio 34/3, 20133 Milano, Italy}
\altaffiliation{Current address: Institute for Molecular Engineering, University of Chicago, 5801 South Ellis Avenue, Chicago, Illinois 60637, USA}
\author{Cristiana Di Valentin}
\affiliation{Dipartimento di Scienza dei Materiali, Universit\`a di Milano-Bicocca, via R. Cozzi 55, 20125 Milan, Italy}
\author{Giovanni Onida}
\affiliation{Dipartimento di Fisica dell'Universit\`a degli Studi di Milano, via Celoria 16, 20133 Milano, Italy}
\alsoaffiliation{European Theoretical Spectroscopy Facility (ETSF)}
\author{Carlo Enrico Bottani}
\affiliation{Department of Energy, Politecnico di Milano, via Ponzio 34/3, 20133 Milano, Italy}
\alsoaffiliation{Center for Nano Science and Technology @Polimi, Istituto Italiano di Tecnologia, via Pascoli 70/3, 20133 Milano, Italy}
\author{Gianfranco Pacchioni}
\affiliation{Dipartimento di Scienza dei Materiali, Universit\`a di Milano-Bicocca, via R. Cozzi 55, 20125 Milan, Italy}

\title{Anisotropic Effects of Oxygen Vacancies on Electrochromic Properties and Conductivity of $\gamma$-Monoclinic WO$_3$}

\begin{document}

\begin{abstract}

Tungsten trioxide (WO$_3$) is a paradigmatic electrochromic material, whose peculiar optical properties in the presence of oxygen vacancies or intercalated alkali atoms have been observed and investigated for a long time. In this paper we propose a rationalization of experiments based on first-principles calculations of optical and electrical properties of oxygen deficient (reduced) WO$_3$. Our approach is based on a parameter-free dielectric-dependent hybrid density functional methodology, used in combination with the charge transition levels formalism, for studying excitation mechanisms in the presence of defects. Our results indicate that oxygen vacancies lead to a different physics in $\gamma$-monoclinic WO$_3$, depending on the orientation of the W-O-W chain where the vacancy is created, thus evidencing strong anisotropic effects rooted in the peculiar structural properties of the original nondefective monoclinic cell. Different types of oxygen vacancies can hence be classified on the basis of the calculated ground state properties, electronic structure, and excitation/emission energies, giving a satisfactory explanation to a range of experimental observations made on oxygen deficient WO$_3$.

\end{abstract}

\section{Introduction}
\label{sec:intro}

Electrochromism is the property by which a material changes its optical properties in  response to counterbalanced injection or extraction of charge and as a consequence of redox processes occurring thereby.\cite{mortimer2011} Remarkably, this phenomenon is exhibited by transition metal oxides, where metal cations can readily vary their oxidation state.\cite{granqvist1995} Among metal oxides, tungsten trioxide (WO$_3$) certainly stands out as the most extensively studied electrochromic material. In fact, the first observations of a change in the coloration characteristics of pure WO$_3$ upon reduction in a stream of hydrogen gas date back to 1815.\cite{berzelius1815} Upon hydrogen or alkali metal ions intercalation, WO$_3$ shows a rise in coloration intensity, whose properties depend on impurities type and concentration.\cite{gerard1980,brown1954, dini1996} Oxygen (O) deficiency has also been recognized as responsible for the emergence of electrochromism in undoped WO$_3$, as well as for its enhancement in ion-intercalated samples.\cite{deneuville1978,yoshimura1985,niklasson2007} Devices exploiting the electrochromic effect have been conceived, starting from the pioneering work of Deb,\cite{deb1969} and today include, among others, smart windows and gas sensors.\cite{deb2008}  

A commonly accepted view is that electrochromic properties of alkali doped and substoichiometric WO$_3$ are intimately connected with formation of W$^{5+}$ (former case), or W$^{5+}$ and W$^{4+}$ (latter case) ionic species, arising from the localization of excess electrons introduced by the impurity (alkali atom or O vacancy) on the $5d$ orbitals of the W$^{6+}$ cations of pristine tungsten oxide. Several models have been proposed to rationalize the underlying coloration mechanisms. Quite generally, these models are founded on the following picture: absorption of a photon ($h\nu$) in the material is responsible for charge hopping from one W$^{5+}$ site to the neighboring W$^{6+}$ one (W$^{5+}_1$ + W$^{6+}_2$ + $h\nu$ $\rightarrow$ W$^{6+}_1$ + W$^{5+}_2$), or from one W$^{5+}$ to another W$^{5+}$, turning W$^{5+}$ species into W$^{4+}$ (W$^{5+}_1$ + W$^{5+}_2$ + $h\nu$ $\rightarrow$ W$^{6+}_1$ + W$^{4+}_2$). The latter is often referred to in the literature as the intervalence charge transfer model.\cite{faughnan1975, granqvist1995} A more realistic description should explicitly take into account formation of polarons as a result of electron localization on $5d$ orbitals of W atoms in the presence of strong electron-phonon coupling, which is often present in transition metal oxides.\cite{austin1969} In this framework, coloration may be explained in terms of photon assisted charge hopping accompanied by a large lattice distortion (polaron hopping) between different W sites.\cite{schirmer1977, salje1978} From a quite different perspective, Deb attempted to explain optical absorption and electrical conductivity measurements on undoped WO$_3$ thin films in terms of formation of color centers, which he argued to be due to positively charged O vacancies.\cite{deb1973} More recently, the same author revisited the issue suggesting a model in which the O vacancy in different charge states is assumed responsible for the formation of defect states resonant with the valence band (neutral vacancy), with the conduction band (doubly charged vacancy), or localized in the band gap (singly charged vacancy).\cite{deb2008} According to Deb's proposal, the defect state associated to a singly charged vacancy should be responsible for the peculiar optical features observed in nonoxidized WO$_3$ thin films.

The influence of substoichiometry on the electronic structure of WO$_3$ has been widely investigated by means of various spectroscopic techniques: X-ray and UV photoelectron,\cite{brigans1981,hoechst1982,szilagyi2012,vasilopoulou2014} optical absorption,\cite{deneuville1978,johansson2014,hollinger1976,ozkan2003,sato2008,sato2010,vasilopoulou2014,vourdas2012,deb1973} as well as photoluminescence (PL) spectra\cite{paracchini1982, manfredi1981, lee2003,feng2005,szilagyi2012,johansson2014,vasilopoulou2014,park2011,luo2007,lu2006,wang2009,niederberger2002,rajagopal2009} have been reported in the literature for both amorphous and crystalline tungsten oxide. Most of these studies essentially agree on recognizing the role of O vacancies as the origin of defect states in the band gap of WO$_3$, acting as optical centers involved in sub-gap electronic transitions. However, an ultimate and unambiguous explanation of the fundamental excitation and decay mechanisms has not yet been accomplished; for example, the origin of two emission peaks observed in the blue and UV-vis regions of PL spectra have been interchangeably assigned to O vacancies and band-to-band transitions in different experimental works.\cite{luo2007,lee2003,feng2005} Apart from interpretative controversies, electrochromic and related properties of WO$_3$ have been reported to be strongly dependent, e.g., on sample crystallinity (or lack thereof), microstructure and defectiveness, which of course are influenced by growth conditions.\cite{ozkan2003,kubo1998} Hence, comparison between different measurements, as well as between experiments and theoretical calculations, should be carried out with great care.

For the sake of simplicity, and because of its relevance to interpretation of room-temperature measurements on this material, we will here limit our attention to the effects of substoichiometry in the monoclinic crystalline phase of WO$_3$ stable in the temperature range from 300 to 603~K (referred to as $\gamma$-WO$_3$).\cite{loopstra1966} $\gamma$-WO$_3$ transforms into the $\beta$ triclinic crystal  structure of WO$_3$ below 300 K,\cite{woodward1995} and there is indication that both phases can be formed in nanocrystalline films grown under certain conditions, thus possibly coexisting at room temperature; however, the contribution from monoclinic $\gamma$-WO$_3$ for substoichiometric films has been reported to be predominant.\cite{johansson2012}

From the theoretical side, density-functional theory (DFT) has been widely applied to the study of O deficient WO$_3$ (WO$_{3-x}$).\cite{chatten2005,lambert2006,wang2011,lambert2012,bondarenko2015} In particular, the hybrid functional DFT calculations of Wang~\emph{et al.} showed that the electronic structure of WO$_{3-x}$ is strongly dependent on the concentration of O vacancies, as well as on their position and orientation in the monoclinic cell;\cite{wang2011} the related optical properties were analyzed based on the computed positions of charge transition levels (CTLs) in the band gap. Bondarenko~\emph{et al.} investigated polaron hopping in O deficient and lithium doped $\gamma$-WO$_3$ using the DFT + $U$ method.\cite{bondarenko2015} Both methodologies have proved to describe in at least a qualitatively correct way ground state properties (in particular, charge localization) and the electronic structure. Calculations on alkali atom intercalated WO$_3$ have also revealed the existence of a robust and profound correlation between the electronic properties and the crystalline structure of this material.\cite{bullet1983,stachiotti1997,stashans1997,walkingshaw2004,tosoni2014}

In this paper, we employ DFT in combination with a self-consistent hybrid exchange-correlation (xc) functional,\cite{skone2014,gerosa2015a} together with the CTLs approach,\cite{vandewalle2004,lany2008} to investigate neutral and charged O vacancies in $\gamma$-WO$_3$. At difference with commonly employed hybrid functionals such as PBE0,\cite{perdew1996,adamo1999} B3LYP,\cite{becke1993,stephens1994} or HSE06,\cite{heyd2003} the xc functional used in the present study features a nonempirical fraction of exact exchange $\alpha$, which is computed from first principles and in a self-consistent fashion by exploiting an approximate but rigorous relationship with the dielectric constant of the material.\cite{skone2014,gerosa2015a} In a recent work, we showed that this computational methodology is suitable for a quantitative analysis of electronic and thermal transitions in O deficient metal oxide materials.\cite{gerosa2015b} The goal of the present investigation is to extend the work reported by some of us in Ref.~\citenum{wang2011} by considering a lower and more realistic concentration of dopants and a more accurate exchange-correlation functional. Using supercell models of unprecedented size (at least 127 atoms), with the aim of minimizing any periodic image defect interaction effects, we propose here for the first time a complete rationalization of the complex experimental scenario concerning optical and electrical properties of substoichiometric $\gamma$-WO$_3$, explaining why O vacancies are at the origin of both deep sub-gap optical transitions and $n$-type conductivity in this material.

\begin{figure}[tb]
 \centering
  \includegraphics{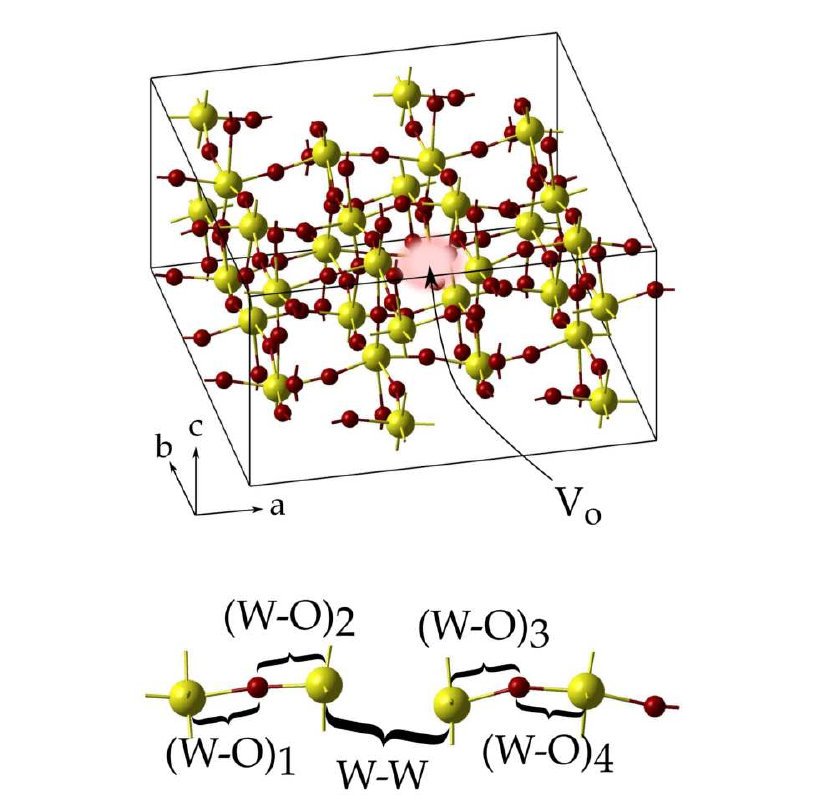}
\caption{Sample simulation cell employed for defect calculations (top), obtained from a bulk 128-atom $\gamma$-WO$_3$ cell by removing an oxygen atom (here a V$_{\text{O},a}$ vacancy is shown as an example). Bond lengths and atomic distances along the defective W-O-W chain (bottom).}
\label{fig:cell}
\end{figure}

\begin{table}[tb]
\caption{\label{tab1} Distance (\AA) between W and O atoms along the W-O-W chain interrupted by a neutral O vacancy, and between W atoms nearest neighbors to the vacancy for the unrelaxed (Unrelax) and the relaxed (Relax) defective structure (see Figure~\ref{fig:cell}). The unrelaxed configuration corresponds to the equilibrium one for the pristine supercell. The formation energy $E_\text f$ (eV) for a neutral O vacancy in the relaxed structure is also reported.}
\begin{tabular}{lcccccc}
\hline
     & \multicolumn{2}{c}{V$_{\text{O},a}$} & \multicolumn{2}{c}{V$_{\text{O},b}$} & \multicolumn{2}{c}{V$_{\text{O},c}$} \\
  & Unrelax & Relax & Unrelax & Relax & Unrelax & Relax \\  
\hline
(W-O)$_1$ & 1.88 & 1.86    & 2.11 & 2.04    & 2.21 & 2.08 \\
(W-O)$_2$ & 1.90 & 1.92    & 1.77 & 1.75    & 1.74 & 1.73 \\
$\,\,$W-W & 3.75 & 3.61    & 3.84 & 4.34    & 3.88 & 4.49 \\
(W-O)$_3$ & 1.88 & 1.83    & 2.11 & 1.83    & 2.21 & 1.74 \\
(W-O)$_4$ & 1.90 & 1.97    & 1.77 & 1.89    & 1.74 & 2.01 \\
E$_\text f$ &    & 5.0     &      & 4.6     &      & 4.2  \\
\hline
\end{tabular}
\end{table}

\section{Computational details}
\label{sec:computational}

DFT calculations were performed within the linear combination of atomic orbitals approach as implemented in the \textsc{crystal09} code.\cite{crystal09-manual,dovesi2005} An effective small core pseudopotential was employed for modeling core electrons in the W atom,\cite{hay1985} while valence electrons were described using the basis sets defined in Ref.~\citenum{wang2011a} ($5p$, $5d$, $6sp$ W electrons in the valence). Instead, the O atom was treated at the all-electron level, using the basis set reported in Ref.~\citenum{ruiz2003}.

A self-consistent hybrid functional was adopted for the treatment of exchange and correlation;\cite{skone2014,gerosa2015a} the exchange fraction $\alpha$ was evaluated self-consistently with the computed average dielectric constant $\epsilon_{\infty}$\cite{ferrero2008} of pristine bulk WO$_3$. The computed value of this quantity, which is 4.56, is in good agreement with the experimental value for $\gamma$-WO$_3$, which is reported to be 4.81;\cite{hutchins2006} the corresponding exchange fraction is given by the inverse of the dielectric constant, and thus amounts at 21.9\%.

The O vacancy was modeled in embedding $\gamma$-monoclinic WO$_3$ supercells comprising at least 127 atoms (see Figure~\ref{fig:cell}); atomic positions and lattice parameters were fully relaxed for the pristine cell using the self-consistent hybrid functional. For the defective supercell, further optimization of the atomic positions was carried out while keeping lattice parameters fixed. Supercells were obtained by expanding the primitive $\gamma$-monoclinic cell (32 atoms) along its crystallographic axes $a$, $b$, and $c$, as detailed in the following:
\begin{itemize}
 \item $2\times2\times1$ expansion, when the vacancy is along the W-O-W chain oriented as the $a$ axis (V$_{\text{O},a}$);
 \item $2\times2\times1$ expansion, when the vacancy is along the W-O-W chain oriented as the $b$ axis (V$_{\text{O},b}$);
 \item $2\times1\times2$ expansion, when the vacancy is along the W-O-W chain oriented as the $c$ axis (V$_{\text{O},c}$).
\end{itemize}

Optical CTLs were computed on the basis of the formalism illustrated in Refs.~\citenum{vandewalle2004} and \citenum{lany2008}. In particular, total energy differences relative to defect charge state variations were expressed in terms of defect Kohn-Sham (KS) eigenvalues evaluated at the $\Gamma$ point, following the approach proposed in Refs.~\citenum{gallino2010} and \citenum{alkauskas2007}. Thermodynamic levels were obtained from the corresponding optical ones by adding the relaxation energy computed in the final defect charge state configuration.\cite{gerosa2015b} The $1s$ KS eigenvalue of oxygen was taken as reference for aligning band structures in defect calculations. The spurious electrostatic interaction between charged image defects was accounted for by correcting the KS eigenvalues according to the procedure illustrated in Ref.~\citenum{chen2013} and based on the Makov-Payne correction scheme;\cite{makov1995} the correction formula was further simplified on the basis of the Lany-Zunger approximation.\cite{lany2008}

The O vacancy formation energy was computed with reference to O rich conditions, i.e. the total energy difference between defective and pristine supercells were compared with half the ground state total energy of the O$_2$ molecule (in the triplet configuration), computed using the PBE0 functional.\cite{perdew1996, adamo1999}

\section{Results and Discussion}

\begin{figure}[tb]
\centering
\includegraphics{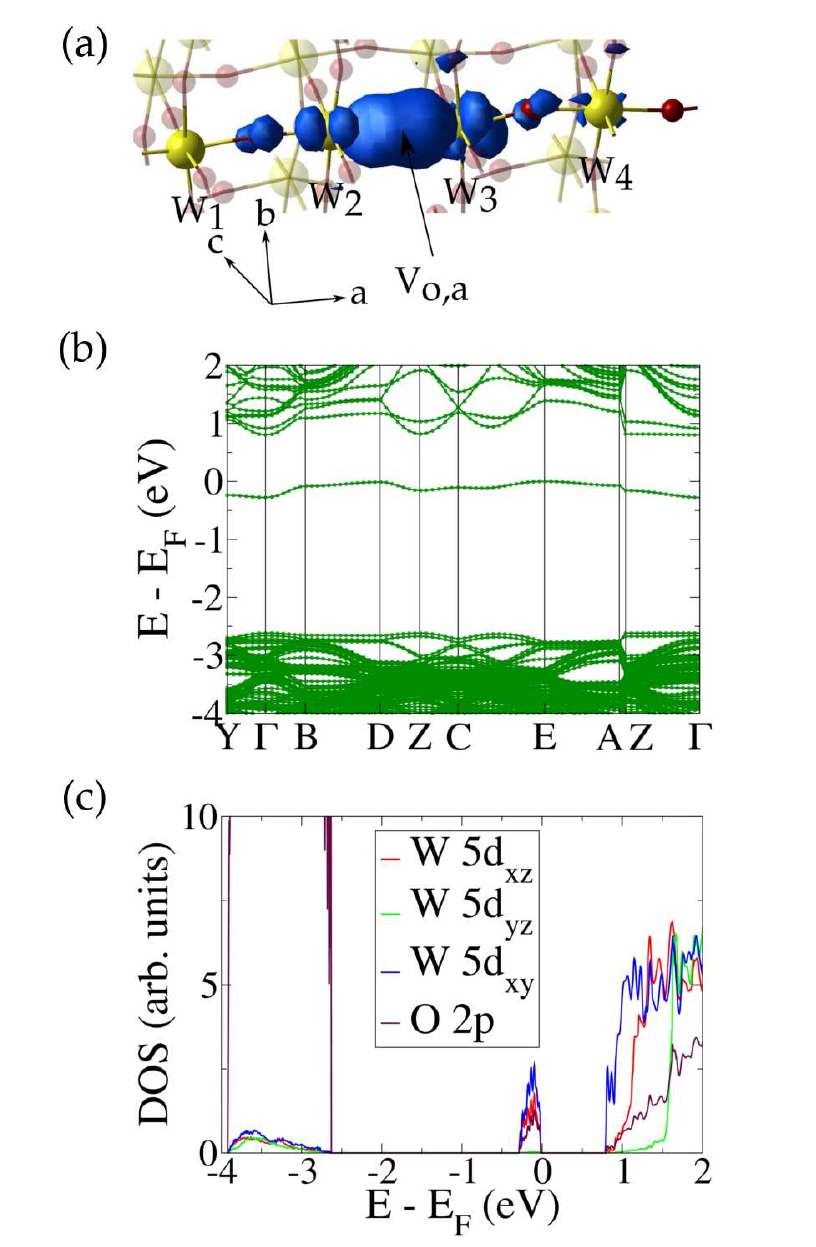}
\caption{O vacancy along the $a$ axis (V$_{\text O,a}$). (a) Isosurface plot of the charge density associated with the excess electrons for a neutral vacancy. The yellow (large) and red (small) spheres represent W and O atoms, respectively. Corresponding (b) electronic band structure, and (c) density of states (DOS) computed at the equilibrium geometry.}
\label{fig:voa}
\end{figure}

\subsection{Ground state and electronic structure of $\gamma$-WO$_{3-x}$: effect of the O vacancy orientation}

A sound description of the ground state and electronic structure of defective materials is an essential starting point for properly understanding excitations. Localization of defect donated charge carriers (excess charge) is often observed in these systems, either as due to lattice distortion and consequent charge trapping (polaron formation) or to a purely electronic effect. In either cases, local or semilocal DFT cannot properly capture such localization, mainly due to the self-interaction error.\cite{divalentin2006} Exact-exchange hybrid functionals and Hubbard-corrected DFT are two viable solutions to such issue, and have been successfully applied to defective transition metal oxides.\cite{bondarenko2015,wang2011,divalentin2014,morgan2009} In particular, we recently demonstrated that the self-consistent hybrid functional employed in this study is able to correctly model O deficient metal oxides\cite{gerosa2015b} and proved to be largely self-interaction free even for a model solid state system for which popular hybrid functionals are not.\cite{gerosa2015c}

The primitive cell of room-temperature monoclinic WO$_3$ is constituted by a three-dimensional network of corner-sharing WO$_6$ octahedra whose axes are tilted relative to each other and where the central W atom exhibits an off-centering displacement along the three crystallographic directions $a$, $b$ and $c$. This determines a deviation from the ideal perovskite ABO$_3$ crystal structure of simple cubic WO$_3$ (A-type atom missing), as a result of a pseudo Jahn-Teller distortion.\cite{bersuker2013} The electronic structure is strongly dependent on the crystal structure, as many previous theoretical calculations have evidenced.\cite{bullet1983,stashans1997,chatten2005,wang2011a,dewijs1999,ping2014,gerosa2015a} Most notably, the band gap drastically increases passing from the most symmetric cubic phase to the $\gamma$-monoclinic structure: this transition has been traced to a change in the hybridization properties of the valence and conduction band edges as a consequence of the W-O bond length splitting (occurring as the W atom moves off the WO$_6$ octahedron center along $a$, $b$, and $c$), as well as to tilting of the octahedra.\cite{ping2014,dewijs1999}
The bond length splitting shows up as an alternation of long and short W-O bonds along the W-O-W atomic chains; in bulk $\gamma$-WO$_3$, its magnitude is more pronounced for the chains oriented as the $b$ and $c$ axes (see Table~\ref{tab1}). This structural anisotropy reflects onto a different behavior of the O vacancy depending on the direction of the W-O-W chain where it is created.\cite{bondarenko2015,wang2011,lambert2006} The electrochromic properties are also expected to be affected by such anisotropy, as will be discussed in the next section.

\begin{figure}[tb]
\centering
\includegraphics{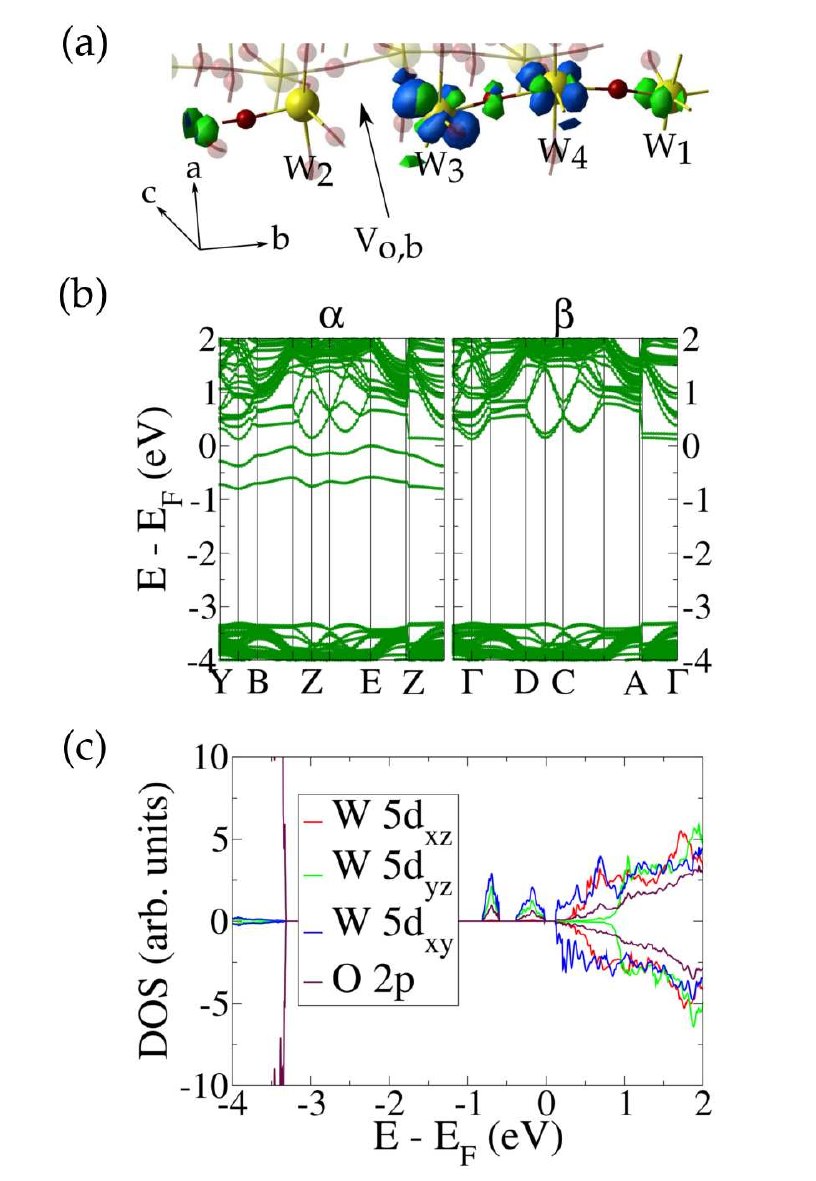}
\caption{O vacancy along the $b$ axis (V$_{\text O,b}$). (a) Isosurface plot of the charge density associated with the excess electrons for a neutral vacancy. The yellow (large) and red (small) spheres represent W and O atoms, respectively. The blue (green) component of the charge density is relative to the electron occupying the deepest (shallowest) defect level. Corresponding (b) electronic band structure, and (c) density of states (DOS) computed at the equilibrium geometry. }
\label{fig:vob}
\end{figure}

From a DFT perspective, one may classify the nature of O vacancies on the basis of different exhibited features: (i) the density distribution of the excess charge in the ground state, and the induced structural relaxation around the defect; (ii) the magnetic or nonmagnetic nature of the ground state; (iii) the energy required to create an O vacancy in the pristine material (formation energy, $E_{\text f}$); (iv) the electronic properties of the defect states formed in the band gap. As we observe below, these features are in fact largely interconnected. Clearly, results are affected by the atomistic model employed to study the defective system; since we ideally aim at understanding the nature of isolated defects in an otherwise pristine crystal, defect-defect quantum-mechanical, electrostatic, and elastic interactions should be minimized.\cite{freysoldt2014} To this purpose, employed supercells were obtained by doubling the bulk primitive $\gamma$-WO$_3$ cell along the direction of the W-O-W chain interrupted by a vacancy. As evidenced in preliminary calculations, doubling the cell axis along $a$ in the V$_{\text{O},b}$ and  V$_{\text{O},c}$ cases is also necessary to reduce the overlapping of defect wavefunctions along this direction;\footnote{Such quantum-mechanical defect-defect interaction appears in the computed band structure as a large dispersion of the corresponding defect states, which partially merge with the conduction band, leading to a metallic system. At the hybrid DFT level, similar results were obtained in the previous study of Wang~\emph{et al.}, which evidenced the dependence of electronic properties on O vacancy concentration.\cite{wang2011}} taking this fact into account, the core set of our calculations employs supercells of 127 atoms.

A common feature characterizing the electronic structure of all the models of $\gamma$-WO$_{3-x}$ investigated here is the appearance of one or more defect states (according to the magnetic nature of the ground state) in the range $0-1$~eV below the conduction band (CB). The band gap remains essentially unaltered with respect to that of stoichiometric $\gamma$-WO$_{3}$, which is computed to be $3.45$~eV. This value is in good agreement with the one measured by means of direct and inverse photoemission ($3.38\pm0.2$~eV, Ref.~\citenum{meyer2010}; $3.39\pm0.2$~eV, Ref.~\citenum{kroger2009}) which has been interpreted as representative of the bulk, rather than the surface, band gap.\cite{ping2014} The optical gap is substantially smaller, being measured in the broad range from 2.6 to 2.9~eV,\cite{kleperis1997,koffyberg1979,salje1974,hardee1977,davazoglou1988,berak1970} and has been shown to be reproduced by first-principles calculations once the spin-orbit interaction and electron-phonon (finite temperature) renormalization effects are included.\cite{ping2013}

\begin{figure}[!h]
\centering
\includegraphics{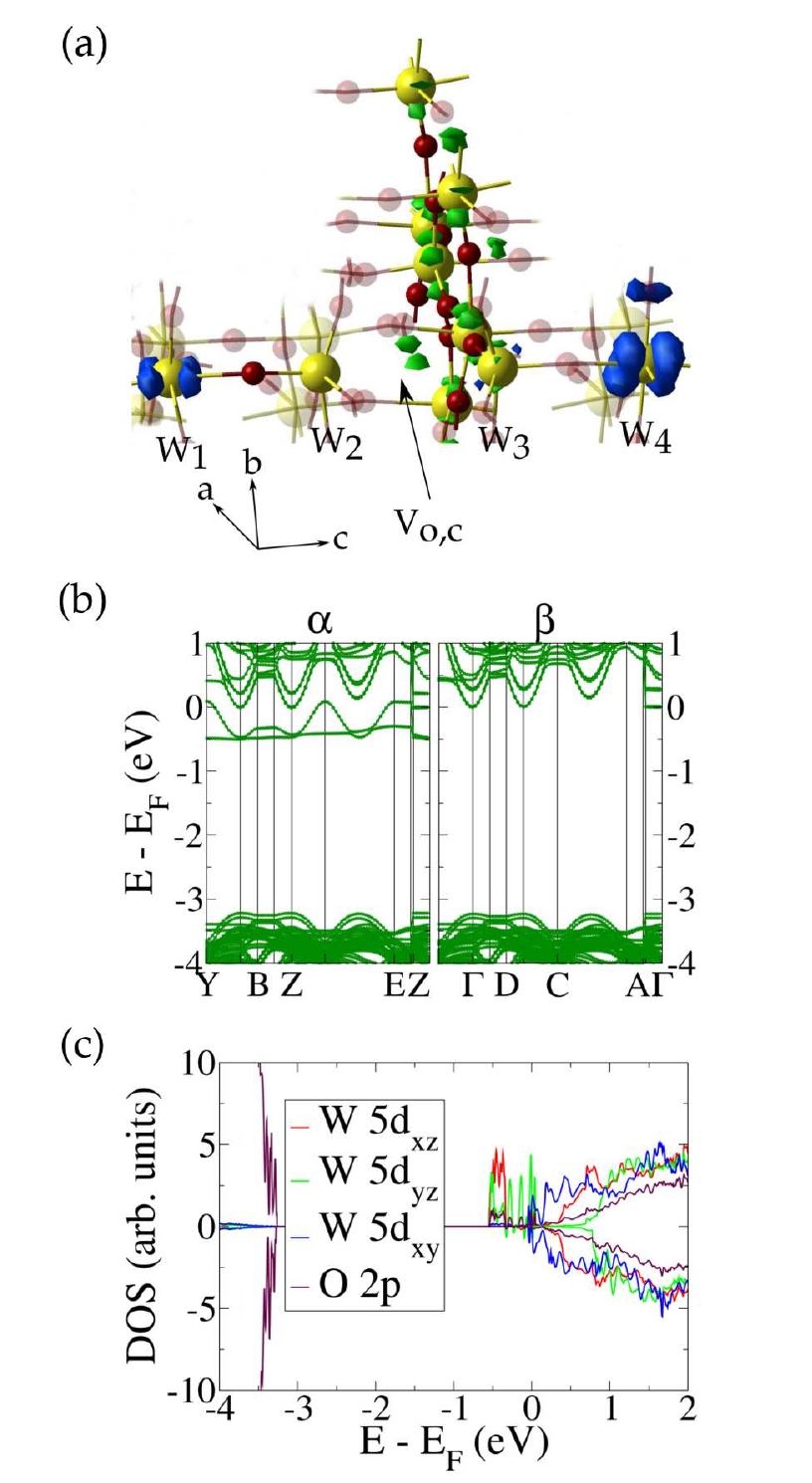}
\caption{O vacancy along the $c$ axis (V$_{\text O,c}$). (a) Isosurface plot of the charge density associated with the excess electrons for a neutral vacancy. The yellow (large) and red (small) spheres represent W and O atoms, respectively. The blue (green) component of the charge density is relative to the electron occupying the flat (dispersive) defect level. Corresponding (b) electronic band structure, and (c) density of states (DOS) computed at the equilibrium geometry.}
\label{fig:voc}
\end{figure}

\begin{table}[tb]
\caption{\label{tab2} Spin density at different W sites along the defective W-O-W chains for the singly charged V$_{\text{O},a}$ (at the neutral geometry), and for both neutral and singly charged V$_{\text{O},b}$ and V$_{\text{O},c}$ vacancies (at the respective relaxed geometry). Neutral and singly charged O vacancies are indicated as  V$_{\text{O}}^{0}$ and V$_{\text{O}}^{+1}$, respectively. W atoms are numbered according to Figures~\ref{fig:voa}(a), \ref{fig:vob}(a), and \ref{fig:voc}(a), respectively.}
\centering
\begin{tabular}{lcccc}
\hline
Type of vacancy & W$_1$  & W$_2$ & W$_3$ & W$_4$ \\
\hline
V$_{\text{O},a}^{+1}$@V$_{\text{O},a}^{0}$ geometry & 0.02 & 0.60 & 0.21 & $<$0.01 \\
\medskip
V$_{\text{O},b}^{0}$  & 0.35 & 0.03 & 0.77 & 0.40 \\
V$_{\text{O},b}^{+1}$ & 0.02 & 0.03 & 0.40 & 0.34 \\
\medskip
V$_{\text{O},c}^{0}$  & 0.54 & 0.06 & 0.19 & 0.88 \\
V$_{\text{O},c}^{+1}$ & 0.59 & 0.08 & $<$0.01 & 0.18 \\
\hline
\end{tabular}
\end{table}

In the W-O-W chain along the $a$ axis, W-O bond length splitting is only slightly developed, as shown in Table~\ref{tab1}. This feature seems essential to prevent large structural distortions to occur as the system reaches its equilibrium configuration in the presence of a neutral vacancy. In fact, the relaxed geometry configuration exhibits only small changes in the W-O bond lengths with respect to the unrelaxed one, in contrast with what is observed for the W-O-W chains oriented as $b$ and $c$. The computed ground state for V$_{\text O,a}$ is a closed-shell singlet (the triplet solution is $\sim 0.2$~eV higher in energy), and the excess electrons, which localize in the vacancy void [see Figure~\ref{fig:voa}(a)], pair up in a bonding-like state between undercoordinated W atoms (whose distance thus decreases with respect to the stoichiometric structure). The behavior of V$_{\text O,a}$ is thus closely resembling that of O vacancies in nonreducible transition metal oxides like zirconium dioxide,\cite{gerosa2015b} in which vacancy sites similar to color centers, typically observed in strongly ionic oxide compounds like magnesium oxide, are formed;\cite{pacchioni2015} in fact, the formation of this center requires an energy as large as 5.0~eV under O rich conditions (see Table~\ref{tab1}). Formation of a doubly occupied closed-shell defect state deep in the band gap [see Figure~\ref{fig:voa}(b)], small structural relaxation around the defect (as opposed to creation of large polaron configurations in other oxides), and higher formation energy are all fingerprints of development of color centers in nonreducible oxides which also characterize the V$_{\text O,a}$ center in $\gamma$-WO$_{3}$.

\begin{figure*}[tb]
\centering
\includegraphics{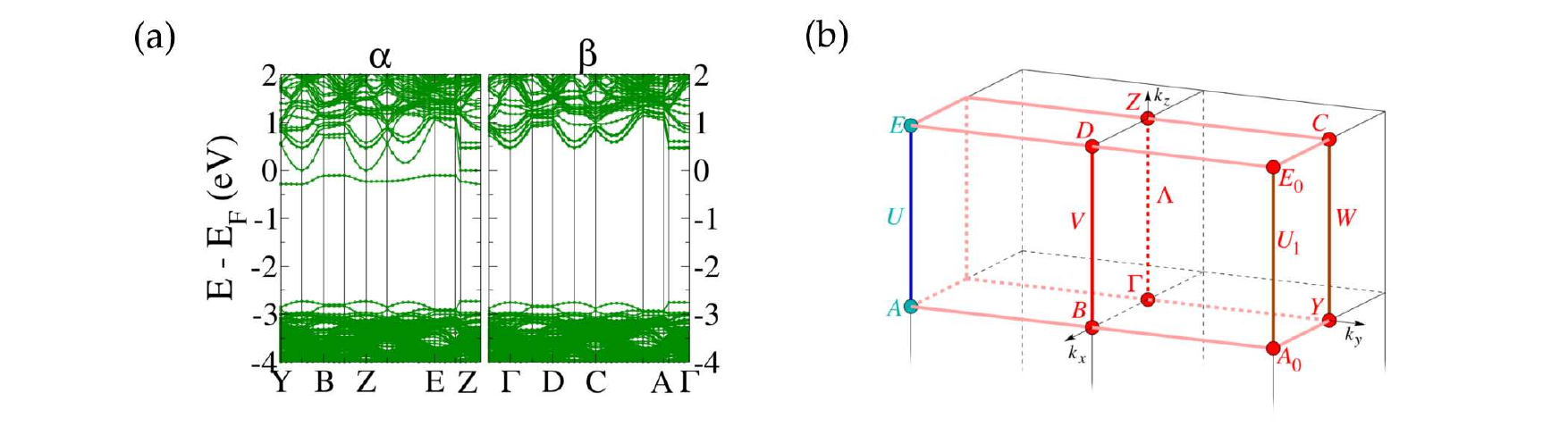}
\caption{(a) Electronic band structure calculated for a neutral V$_{\text O,c}$ vacancy embedded in a large $2\times2\times2$ supercell of 255 atoms. (b) Brillouin zone of $\gamma$-monoclinic WO$_3$: labeling of the high-symmetry $\mathbf{k}$ points and directions (figure adapted from Bilbao Crystallographic Server\cite{aroyo2014}).}
\label{fig:voc2x2x2}
\end{figure*}

When the properties of V$_{\text O,b}$ and V$_{\text O,c}$ are investigated, a remarkably different behavior can be noticed. In fact, the ground state for both vacancy configurations is found to be a triplet, the closed-shell solution being much higher in energy ($\sim 0.8$~eV for V$_{\text O,b}$ and $\sim 1.1$~eV for V$_{\text O,c}$). The triplet ground state is characterized by peculiar features, which mark the difference with the V$_{\text O,a}$ case. In fact, the excess electrons are now localized on all the W atoms belonging to the chain interrupted by a vacancy, with a substantial contribution from the fully coordinated ones (see Figures~\ref{fig:vob}(a) and \ref{fig:voc}(a), and Table~\ref{tab2}). Such charge redistribution comes with a considerable structural distortion around the defect, which makes the undercoordinated W atoms move apart by as much as 0.5~$\text{\AA}$ (V$_{\text O,b}$) and 0.7~$\text{\AA}$ (V$_{\text O,c}$) with respect to the nondefective structure. The strong coupling between charge localization and structural relaxation reflects the polaronic nature of these defects. The same phenomenology is encountered in reducible oxide materials like titanium dioxide.\cite{gerosa2015b,deak2012} Notice that, as expected, this corresponds to a lower O vacancy formation energy for V$_{\text O,b}$ and V$_{\text O,c}$ with respect to V$_{\text O,a}$, as reported in Table~\ref{tab1}. Results of previous calculations using supercells of at least 64 atoms are consistent with our finding that $E_\text f[\text V_{\text O,a}]>E_\text f[\text V_{\text O,b}]>E_\text f[\text V_{\text O,c}]$.\cite{bondarenko2015,lambert2006} Hence, assuming a Boltzmann distribution of O vacancies in the material, V$_{\text O,c}$ centers are expected to exhibit a larger concentration, as they are found to be more stable with respect to V$_{\text O,b}$ and V$_{\text O,a}$ by 0.4~eV and 0.8~eV, respectively; one should of course take account of the predominance of V$_{\text O,c}$ vacancies when interpreting results related to optical and electrical properties.

In comparison with V$_{\text O,a}$, the gap states introduced by V$_{\text O,b}$ and V$_{\text O,c}$ have overall shallower character, spanning a region from 0 to 1~eV below the CB. In particular, in the case of V$_{\text O,c}$ vacancies, one of these states clearly follows the dispersion behavior of the CB edge [see Figure~\ref{fig:voc}(b)]. The electron occupying this state partially delocalizes over the plane containing the $a$ and $b$ axes and perpendicular to $c$, as represented by the green component of the isosurface charge density shown in Figure~\ref{fig:voc}(a). One may wonder whether this picture has a genuine physical basis or should instead be ascribed to a residual interaction between image defects in neighboring cells. In order to rule out the latter possibility, we further investigated the electronic structure of the neutral V$_{\text O,c}$ center using a supercell with a doubled $b$ lattice constant (255 atoms). As shown in Figure~\ref{fig:voc2x2x2}(a), the properties of the two gap states are qualitatively the same as those calculated in the smaller supercell. The flat level retains the nature of a defect state, being positioned $0.5-0.7$~eV below the CB, whereas the dispersive level has a clear conduction state character, which emerges even more apparently in this case. In particular, the parabolic dispersion along the Y-$\Gamma$-B and the D-Z-C Brillouin zone lines lying in the plane perpendicular to $c$ [see Figure~\ref{fig:voc2x2x2}(b)] suggests a free-electron-like behavior for the excess electron occupying this level. This in turn suggests that the neutral V$_{\text O,c}$ vacancy is unstable against transfer of an excess electron to the CB, making the singly charged vacancy the stable configuration for V$_{\text O,c}$ in the absence of external optical excitations. It is thus explained already in the DFT framework the experimental observation of O vacancies being at the origin of $n$-type band conductivity in substoichiometric WO$_3$.\cite{gillet2003,gillet2004,berak1970,aguir2002} The issue of how and what types of O vacancies can also give rise to deep optical transitions will be dealt with in the following section.

Analysis of the electronic density of states (DOS), shown in Figures~\ref{fig:voa}(c), \ref{fig:vob}(c), and \ref{fig:voc}(c), reveals that, as expected, the valence and conduction band edges retain the same character as in stoichiometric WO$_3$:\cite{wang2011} the top of the valence band (VB) mainly comes from $2p$ states of O atoms, while the bottom of the CB is made up of W $5d$ states, with a major contribution from the $d_{xy}$ orbital ($d_{xz}$ and $d_{yz}$ orbitals begin to play a role at higher energy). Instead, defect states in the gap possess a different character depending on the O vacancy orientation.\footnote{See the Supporting Information.} In all cases, however, the main contribution is from W $5d$ orbitals, although the O $2p$ contribution is not negligible. 

Singly and doubly charged O vacancies were finally considered, with the ultimate goal of calculating CTLs (see the next section). When passing from a neutral to a singly charged vacancy, the defect state closest to the CB (being in fact a conduction state for V$_{\text O,c}$) is emptied and, upon lattice relaxation in the new charged state, merges with the CB. Thus, only a single unpaired electron remains in the system, and the ground state is now a doublet. Analysis of the spin population of the relevant W $5d$ orbitals along the defective W-O-W chain (Table~\ref{tab2}) allows one to understand quantitatively how excess electrons distribute in the vacancy neighborhood and to what extent this modifies the oxidation state of the involved W ions. In particular, it is observed that the excess electron localizes at multiple W sites along the defective W-O-W chain for singly charged V$_{\text O,b}$ and V$_{\text O,c}$ vacancies, as opposed to the naive picture often implied in polaron absorption models where the electron is assumed to be trapped at a single W site in the immediate proximity of the vacancy (thus reducing the W oxidation number from +6 to +5). On the contrary, for the neutral V$_{\text O,a}$, both excess electrons gets trapped in the vacancy void, even if a contribution from $5d$ orbitals of the surrounding W atoms is still present (see the spin population analysis for a singly charged V$_{\text O,a}$ computed at the neutral vacancy geometry, Table~\ref{tab2}).

\subsection{Interpretation of electrical and optical properties of $\gamma$-WO$_{3-x}$ based on CTLs}
\label{sec:ctl}

Electronic transitions involving defect electrons introduced by O vacancies in the material lead to variations of the O vacancy charge state, and can be linked to optical processes probed in absorption and PL spectroscopy.\cite{gallino2010} Hence, the electrical and optical properties of O deficient $\gamma$-WO$_3$ can be analyzed on the basis of transition energies associated to thermodynamic (adiabatic) and optical (vertical) CTLs, respectively.

\begin{figure}[tb]
\centering
\includegraphics{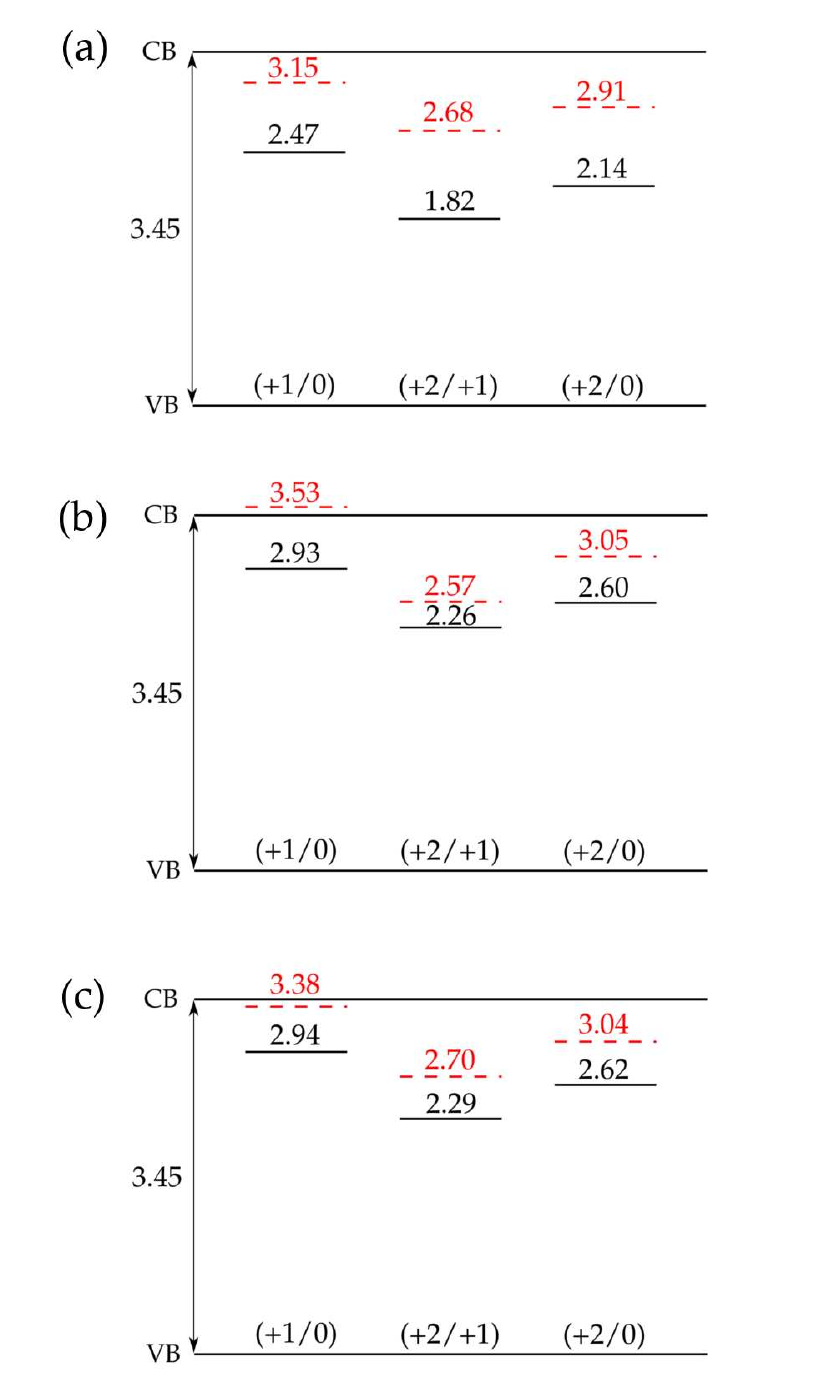}
\caption{Optical (solid, black) and thermodynamic (dashed, red) CTLs, computed for an O vacancy oriented as (a) $a$, (b) $b$, and (c) $c$ crystallographic axes of $\gamma$-WO$_3$. Positions (eV) are given with respect to the top of the VB.}
\label{fig:ctl}
\end{figure}

Figure~\ref{fig:ctl} reports computed CTL positions for all the possible defect related electronic transitions and for differently oriented O vacancies. Firstly, it is noticed that CTLs are generally shallower for V$_{\text O,b}$ and V$_{\text O,c}$ than for V$_{\text O,a}$. More specifically, the $(+1/0)$ thermodynamic level for V$_{\text O,b}$ is positioned above the CB minimum; hence, thermal energy fluctuations at finite temperature can promote one of its defect electron into the CB. For the more abundant V$_{\text O,c}$ vacancy, one excess electron is predicted to be delocalized in the host crystal even when considering a neutral vacancy, as discussed above.\footnote{Since for the neutral V$_{\text O,c}$ vacancy the shallowest defect state is in fact a delocalized conduction state, calculation of the $(+1/0)$ CTLs is of limited significance; instead,  the $(+2/+1)$ CTLs are well defined since the gap state for the singly charged V$_{\text O,c}$ is a well localized defect level.} In contrast, the $(+2/+1)$ thermodynamic level for both V$_{\text O,b}$ and V$_{\text O,c}$ is located deeper in the band gap ($\sim0.9$~eV and 0.75~eV below the CB minimum, respectively). Hence, analysis of charge state stability indicates that, for both vacancy types, the defect center should be singly ionized, thus suggesting that the O deficiency may be at the origin of a finite concentration of conduction electrons in undoped substoichiometric $\gamma$-WO$_3$. The presence of O vacancies hence effectively leads to a semiconductor-metal transition which was already described in the previous theoretical investigation of Wang~\emph{et al.}\cite{wang2011} Such interpretation is corroborated by the experimental evidence of $n$-type conductivity in $\gamma$-monoclinic nanocrystalline WO$_3$ films grown under different reducing conditions.\cite{gillet2003,gillet2004,berak1970,aguir2002,vemuri2010} Amorphous WO$_3$ films have also been reported to exhibit metallic character in the presence of O vacancies.\cite{sato2010,vourdas2012} Theory and experiment thus agree on recognizing the central role played by these defects in the determination of the electrical properties of tungsten trioxide.

The previous analysis also helps identify charge transition mechanisms most likely contributing to characteristic optical features that have been observed in several spectroscopic studies. In the case of V$_{\text O,a}$, for which the stable charge state is the neutral one, all the vertical transitions reported in Figure~\ref{fig:ctl} are virtually possible. However, for the two other O vacancy orientations, for which instead a singly charged state is favored at room temperature, only the $(+2/+1)$ vertical transition is expected to give a substantial contribution.

Intensive experimental work has been carried out in the characterization of the optical properties of substoichiometric tungsten trioxide. Of course, obtained results are sensitive to specific properties of the investigated samples, such as degree of crystallinity, structure at the micro- and nano-scale, and chemical composition. In a recent investigation, Johansson~\emph{et al.} studied $\gamma$-monoclinic nanocrystalline films using PL spectroscopy, performing measurements on samples with a variable concentration of O vacancies.\cite{johansson2014} The obtained PL spectra exhibit several emission peaks, which can be interpreted in light of the computed CTLs. In the most substoichiometric samples, a peak at 537~nm (2.31~eV) was observed which can be explained as due to a $(+2/+1)$ vertical transition involving V$_{\text O,b}$ and V$_{\text O,c}$ vacancies; the relevant CTLs, positioned at 2.26~eV and 2.29~eV above the VB maximum, suggest the emission mechanism to be related to electron-hole recombination involving the VB and the midgap defect level. Similar mechanisms may give rise to the other PL features, whose positions were reported at 490~nm (2.53~eV) and 600~nm (2.07~eV): in this case a major role should be played by the less abundant V$_{\text O,a}$ vacancies, for which the $(+1/0)$ and $(+2/0)$ optical transition levels are computed at 2.47~eV and 2.14~eV above the VB edge. A further peak at 465~nm (2.68~eV) can be only explained in the framework of our calculations by considering a more complex excitation pathway, in which an electron from the CB is initially trapped at a singly charged V$_{\text O,b}$ or V$_{\text O,c}$ site, forming a neutral vacancy; then, simultaneous recombination of both defect electrons with holes in the VB would lead to emission of a photon with an energy of about 2.6~eV [compare with computed values of the $(+2/0)$ optical levels in Figures~\ref{fig:ctl}(b) and \ref{fig:ctl}(c)]. Johansson~\emph{et al.} alternatively explained this feature in terms of a similar multistep $(+1/+2)$ transition, in which a nonrelaxed singly charged vacancy decays into a doubly charged one.\cite{johansson2014}

\begin{table}[tb]
\caption{\label{tab3} Selected results obtained from photoluminescence (PL) spectroscopy measurements performed on O deficient WO$_3$ samples of different chemical composition and/or micro- and nano-structure. Position of the most relevant observed PL peaks are reported. Results reported in Refs.~\citenum{johansson2014} and \citenum{paracchini1982} constitute the reference for comparison with our calculations (see text).}
\begin{tabular}{lcccccc}
\hline
 Sample composition & PL peak position & Reference \\  
\hline
  Nanocrystalline $\gamma$-WO$_3$ & 600~nm (2.07~eV) & \citenum{johansson2014} \\
                                  & 537~nm (2.31~eV) & \citenum{johansson2014} \\
                                  & 490~nm (2.53~eV) & \citenum{johansson2014} \\
                                  & 465~nm (2.68~eV) & \citenum{johansson2014} \\
  Polycrystalline WO$_3$ & 550~nm (2.26~eV) & \citenum{paracchini1982} \\
  W$_{18}$O$_{49}$ nanorods & 435~nm (2.85~eV) & \citenum{feng2005} \\
  W$_{18}$O$_{49}$ nanorods & 437~nm (2.84~eV) & \citenum{lee2003} \\
  Hexagonal WO$_3$ nanowire clusters & 420~nm (2.94~nm) & \citenum{rajagopal2009} \\ 
  $\gamma$-WO$_3$ nanoparticles & 421~nm (2.94~eV) & \citenum{szilagyi2012} \\
  WO$_{3-x}$ nanowires & 467~nm (2.65~eV) & \citenum{luo2007} \textsuperscript{\emph{a}} \\
\hline
\end{tabular}

\flushleft
\textsuperscript{\emph{a}} The authors attributed this peak to band-band transitions rather than to transitions involving O vacancies.
\end{table}

Several other PL spectroscopy studies support the general conclusion that O vacancies induce formation of midgap states in WO$_3$, although their detailed position in the band gap has been found to vary considerably depending on the properties of the investigated samples (see Table~\ref{tab3}), thus making comparison with our calculations less straightforward. In fact, some authors have proposed that O vacancies may even give rise to defect states within the CB, which would thus be at the origin of PL emission features with an associated energy exceeding the band gap value (i.e., in the range of wavelengths from 350~nm to 400~nm).\cite{park2011,luo2007,lu2006,wang2009,niederberger2002} Such an interpretation may be consistent with the model proposed by Deb\cite{deb2008} (see Introduction), but is not in agreement with the picture suggested by our calculations, since the computed optical CTLs are all located within the energy gap. In a comprehensive experimental investigation on amorphous WO$_3$, Vasilopoulou \emph{et al.}\cite{vasilopoulou2014} instead pointed out that O vacancies may be responsible for the formation of deep (just above the VB) gap states in amorphous WO$_3$; this finding is in agreement with the results of the theoretical study of de~Wijs and de~Grott,\cite{dewijs1999a} in which a clear correlation between the existence of such gap states and the presence of shortened (smaller than 3~\AA, compare with Table~\ref{tab1}) W-W distances has been reported. It may be viewed as an extreme case of the closed-shell V$_{\text O,a}$ configuration considered here for the crystalline material. Hence, the intimate relationship between structural and electronic properties in substoichiometric WO$_3$ is once again confirmed.

Evidence of defect states in the gap is also supported by photoemission\cite{brigans1981,hoechst1982,hollinger1976,vasilopoulou2014} and optical absorption\cite{deneuville1978,johansson2014,ozkan2003,sato2008,sato2010,vasilopoulou2014,vourdas2012,deb1973} spectroscopy experiments on reduced WO$_3$ samples in the crystalline or amorphous phase. In particular, measurements on nanocrystalline $\gamma$-WO$_3$ reported absorption bands with maxima at 0.73~eV and in the range from 0.96 to 1.16~eV, depending on deposition pressure.\cite{johansson2014} The emergence of these features can be understood as due to a $(+1/0)$ excitation of an electron from the defect state to the CB for V$_{\text O,a}$ (calculated transition energy 0.98~eV), or a $(+2/+1)$ excitation for V$_{\text O,b}$ or V$_{\text O,c}$ (calculated transition energies are 1.19~eV and 1.16~eV, respectively). However, on the basis of the calculated vacancy formation energies, we expect the strongest contribution to be given by V$_{\text O,c}$ vacancies. Amorphous samples also feature sub-gap absorption, although within a typically higher energy range  with respect to crystalline ones,\cite{deb1973,sato2008,ozkan2003,deneuville1978} due to widening of the band gap in going from the latter to the former structural phase.\cite{nakamura1981,dewijs1999a,deb1973}

\section{Conclusions}

We have performed a thorough DFT investigation of the effect of O deficiency on several properties of $\gamma$-monoclinic WO$_3$, including charge localization and corresponding structural distortion around the vacancy, electronic structure, electrical conductivity and optical transitions.

A different behavior has been identified for O vacancies, depending on the orientation of the W-O-W chain where they are introduced. Such anisotropy can be related to the corresponding structural anisotropy in the nondefective monoclinic cell: long and short W-O bonds alternate along the W-O-W chain oriented as the $b$ and $c$ axes, while bond lengths along $a$ are virtually the same. This results in a stiffer structure in the $a$ direction with respect to $b$ and $c$, something which affects the properties of O vacancies in the relevantly oriented chain: the possibility of a strong lattice distortion around the defect site makes polaronic configurations favored, in a similar way as in reducible transition metal oxides, with an evident impact on the related ground state and electronic properties.

In fact, O vacancies oriented as the $a$ crystallographic axis (V$_{\text O,a}$) tend to form color center configurations, with the excess charge localized in the vacancy void to yield a closed-shell singlet ground state (for the neutral vacancy), and a minimal lattice distortion in its vicinity. Associated CTLs are found rather deep in the gap, suggesting the stable O vacancy charge state to be the neutral one.

On the contrary, O vacancies oriented as the $b$ (V$_{\text O,b}$) and $c$ (V$_{\text O,c}$) axes give rise to polaronic defect centers as a consequence of a strong lattice relaxation occurring along the defective W-O-W chain. The excess charge primarily localizes on the $5d$ orbitals of the W atoms along the chain, yielding, for a neutral vacancy, an open-shell triplet ground state. Computed CTLs are generally shallower than in the V$_{\text O,a}$ case, suggesting singly charged vacancies to be more stable at finite temperature; this explains why experiments measure a finite concentration of conduction electrons in O deficient WO$_3$, making it an $n$-type conductor. Moreover, the formation of V$_{\text O,c}$ vacancies has been found to be favored, due to the lower formation energy, with respect to V$_{\text O,a}$ and V$_{\text O,b}$. For the neutral V$_{\text O,c}$ vacancy, the DFT electronic structure features a delocalized band occupied by an excess electron which is virtually belonging to the conduction band. Hence, for this specific type of vacancy, the connection between substoichiometry and conductivity is already explained by standard band structure calculations.

Notice that the size and shape of the employed model supercell crucially determines the possibility of observing the complex phenomenology described above: in the previous work of some of us,\cite{wang2011} in which much smaller cells of 64 and 32 atoms were considered, the excess charge was in fact inevitably constrained in a limited region of space surrounding the vacancy. As a result, the structural relaxation evidenced in the present investigation could not be captured, and consequently the formation energies for the three differently oriented O vacancies were found to be essentially the same, at variance with what is found when a more realistic structural relaxation and concurring charge localization are accounted for. Interestingly, however, both studies do agree on the magnetic nature of the most stable ground state for V$_{\text O,a}$ (closed-shell singlet), and V$_{\text O,b}$ and V$_{\text O,c}$ (triplet). This indicates that the amount of exact exchange in the hybrid functional used in Ref.~\citenum{wang2011} was indeed sufficient to obtain the correct (with respect to the self-interaction error issue) spin polarized ground state. However, the O vacancy gap states computed for V$_{\text O,b}$ and V$_{\text O,c}$ showed different dispersion features from ours, due to the higher concentration of defects along those crystallographic directions, which especially affects the electronic properties of large polarons. In contrast, the excess charge for V$_{\text O,a}$ tends to remain trapped in the defect site, and hence a fine description of the lattice distortion in its vicinity becomes less important. As a consequence, our results for the electronic structure of V$_{\text O,a}$ closely agree with those reported in Ref.~\citenum{wang2011}; the quantitative differences concerning the computed CTLs are due to the the higher amount of exact exchange built in the self-consistent hybrid functional used here, which already influences the band gap value calculated for the pristine material.

Having achieved an accurate description of the electronic structure, and on the basis of the related computed optical CTLs, photoluminescence and optical absorption spectra have been interpreted in terms of a complex interplay of vertical excitations at inequivalent vacancy sites, providing an overall satisfactorily clear picture of the optical mechanisms occurring in O deficient WO$_3$.

In conclusion, electrochromic properties and conductivity of this technologically important material have been rationalized by means of a nonempirical DFT methodology able to give quantitative insights into a wide spectrum of properties relevant to defective materials. In particular, the attention has been focused on how structurally inequivalent O vacancies give rise to a different phenomenology which may be described as typical of nonreducible or reducible transition metal oxides.

\section{Supporting Information Available}

A detailed discussion of the electronic density of states for V$_{\text O,a}$, V$_{\text O,b}$, and  V$_{\text O,c}$ is reported. The isosurface plot of the charge density associated to excess electrons for V$_{\text O,c}$, as computed using a 255-atom supercell, is shown.

\begin{acknowledgement}

This work has been supported by the Italian MIUR through the FIRB Project RBAP115AYN ``Oxides at the nanoscale: multifunctionality and applications''.
The support of the COST Action CM1104 ``Reducible oxide chemistry, structure and functions'' is also gratefully acknowledged. C.DV. and G.P. acknowledge Cariplo Foundation  for the Grant n. [2013-0615]. G.O. acknowledges
the ETSF-Italy\cite{etsf} for computational support. G.O. and M.G. acknowledge useful discussions with Nicola Manini. M.G. thanks Elisa Albanese for help in running calculations and for useful interactions.

\end{acknowledgement}

\bibliography{manuscript}

\end{document}